\documentclass[12pt]{iopart}

\usepackage{setstack}
\usepackage{iopams}
\usepackage{graphicx}

\newcommand{\ket}[1]{|\!#1\rangle}
\newcommand{\bra}[1]{\langle #1|}
\newcommand{\bess}[1]{{J_{#1}}}

\begin{document}

\title{Ion trap quantum gates with amplitude-modulated laser beams}

\author{Christian F. Roos$^{1,2}$}

\address{$^1$Institut f\"ur Experimentalphysik, Universit\"at Innsbruck,
Technikerstr. 25, A-6020 Innsbruck, Austria}
\address{$^2$Institut f\"ur Quantenoptik und Quanteninformation der
\"Osterreichischen Akademie der Wissenschaften,
Otto-Hittmair-Platz 1, A-6020 Innsbruck, Austria}

\begin{abstract}
In ion traps, entangling gate operations can be realized by a
bichromatic pair of laser beams that collectively interact with
the ions. In this paper, a new method of modelling the laser-ion
interaction is introduced that turns out to be superior to
standard techniques for the description of gate operations on
optical qubits. The treatment allows for a comparison of the
performance of gates based on $\sigma_z\otimes\sigma_z$ and
$\sigma_\phi\otimes\sigma_\phi$
interactions on optical transitions where the bichromatic laser
field can be realized by an amplitude-modulated laser resonant
with the qubit transition. Shaping the amplitude of the
bichromatic laser pulse is shown to make the gates more robust
against experimental imperfections.
\end{abstract}

\section{Introduction}
The processing of information based on the laws of quantum physics
\cite{NielsenChuang00} has become a very active field of research
during the last decade. For the experimental demonstration of
fundamental key results of quantum information theory, ion-trap
based systems have played a major role. The success of ion trap
experiments can be attributed to the fact that encoding quantum
information in either hyperfine ground or meta-stable excited
atomic states provides well-defined quantum bits (qubits) with
long coherence times. The use of lasers for manipulating the qubit
state allows for precisely switchable interactions with low
decoherence rates. The fundamental operations of qubit
initialization, arbitrary single qubit manipulation and quantum
state-detection have already been used in atomic clocks with
single ions for many years. Contrary to other realizations
\cite{NielsenChuang00} of quantum information processing, the most
demanding operation in ion traps consists in the realization of an
entangling gate operation. Because of the repulsive Coulomb force,
the inter-ion distance is orders of magnitude bigger than the
characteristic length scale of any state-dependent interaction
between ions in ground or low-lying excited states. In all current
experiments creating entangled ions
\cite{Schmidt-Kaler03,Leibfried03,Haljan05,Home06}, gate
operations rely on interactions that are mediated by the
vibrational degrees of freedom of the ion string. These gate
operations fall into two categories:
\begin{enumerate}
\item Quantum gates induced by a laser beam that interacts with a
single ion at a time as originally proposed in the seminal paper
by I. Cirac and P. Zoller \cite{CiracZoller95} and later realized
by the Innsbruck ion trapping group \cite{Schmidt-Kaler03}. In
these gates, a single ion is entangled with a vibrational mode
\cite{Monroe95} of the ion string and the entanglement is
subsequently transferred from the vibrational mode to the internal
state of a second ion.
\item Quantum gates induced by a bichromatic laser that
collectively interacts with two or more ions. Here, a vibrational
mode becomes transiently entangled with the qubits before getting
disentangled at the end of the gate operation, resulting in an
effective interactions between the qubits capable of entangling
them. Gates of this type were first proposed by G.~Milburn
\cite{Milburn99,Milburn00}, A.~S{\o}rensen, K.~M{\o}lmer
\cite{SorensenMolmer99,SorensenMolmer00} and E.~Solano
\cite{Solano99}, and subsequently realized by ion trapping groups
in Boulder, Ann Arbor and Oxford
\cite{Leibfried03,Haljan05,Home06}.
\end{enumerate}
Even though both classes of gates are applicable to hyperfine
qubits as well as optical qubits (i.e. qubits encoded in hyperfine
states or in states linked by a dipole-forbidden transition with
an optical wavelength), current experiments with optical qubits
have relied on the former and experiments with hyperfine qubits on
the latter type of interaction. In any case, the main goal
consists in demonstrating fast operations creating entanglement
with high fidelity.

The purpose of the present paper is to discuss bichromatic gate
operations with a focus on implementations using an optical
transition. It turns out that for optical transitions, gate
operations are achievable by illuminating the ions with an
amplitude-modulated laser beam that is resonant with the qubit
transition. The paper is organized as follows: section
\ref{sec:BichromaticGates} reviews different methods of realizing
bichromatic quantum gates and discusses properties that are
specific to their application to optical qubits. In section
\ref{sec:EffectiveHamiltonian}, an effective Hamiltonian for the
laser-ion interaction will be derived by going into a reference
frame rotating at non-uniform speed in order to eliminate
non-resonant excitations of the qubit transition that do not
couple to the vibrational mode. In this way, it will be shown that
for a single ion qubit the interaction is well described by a
Hamiltonian $H=i\hbar(\gamma(t)a^\dagger-\gamma(t)^\ast
a)\sigma_\psi$ where the coupling strength $\gamma$ is
proportional to the laser intensity in the limit of low
intensities but starts to saturate at higher intensities and where
$\sigma_\psi=\vec{\sigma}\cdot \vec{n}_\psi$ is a component of the
Pauli spin operator $\vec{\sigma}$ coupling to a vibrational mode
of the ion described by creation and annihilation operators
$a^\dagger$, $a$. Furthermore, it will be shown that
$\vec{n}_\psi$ depends not only on the particular type of gate
operation but also on the laser intensity and the relative phase
between the two frequencies of the bichromatic field. Equations
(\ref{Heffzz}), (\ref{HeffMS}), (\ref{MSPropagator2}) describing
the action of the gates based on $\sigma_z\otimes\sigma_z$ and on
$\sigma_\phi\otimes\sigma_\phi$ (M{\o}lmer-S{\o}rensen)
interactions are the key results of the paper. For the
M{\o}lmer-S{\o}rensen gate, the result will be compared to the
analysis presented in \cite{SorensenMolmer00}. In addition, the
performance of $\sigma_z\otimes\sigma_z$ and
$\sigma_\phi\otimes\sigma_\phi$ gates will be compared. Section
\ref{sec:ShapedPulses} shows how to use pulse-shaping of the laser
intensity as well as spin echo techniques to make the gates more
robust against fluctuations of the control parameters.

\section{Quantum gate operations based on bichromatic laser fields \label{sec:BichromaticGates}}
\subsection{Driven quantum harmonic oscillator \label{subsec:DrivenOsc}}
The Hamiltonian
$\tilde{H}=\hbar\nu a^\dagger a + \hbar\Omega i(a^\dagger
e^{i\omega t}-a e^{-i\omega t})$
describes a harmonic oscillator oscillating at frequency~$\nu$ and
driven by a force with frequency~$\omega$ and coupling
strength~$\Omega$. Going into an interaction picture defined by
$H_0=\hbar\nu a^\dagger a$ yields the Hamiltonian
$H=\hbar\Omega i(a^\dagger e^{i\delta t} - a e^{-i\delta t}),$
where $\delta=\omega-\nu$. Under the action of the  driving force,
an oscillator that is initially in a coherent state remains in a
coherent state. For a force that is slightly detuned from
resonance, the coherent state maps out a circle in phase space and
returns to the initial state after a period $\tau=2\pi/\delta$.
This operation multiplies the oscillator state by a phase factor
whose magnitude is given by the ratio of the strength of the force
and the detuning as shown in \cite{Leibfried03}.

In order to allow for variations of the driving field's strength,
we  generalize the Hamiltonian to
$H=i\hbar(\gamma(t)a^\dagger-\gamma^\ast(t)a)$ and calculate its
propagator $U(t)$ by using the Baker-Campbell-Hausdorff relation
$\hat{D}(\alpha)\hat{D}(\beta)=\hat{D}(\alpha+\beta)\exp(i\mbox{Im}(\alpha\beta^\ast))$
for the displacement operator $\hat{D}(\alpha)=e^{\alpha
a^\dagger-\alpha^\ast a}$. For the propagator, we find
\begin{equation}
U(t)
=\lim_{n\rightarrow\infty}\prod_{k=1}^n
\exp({-\frac{i}{\hbar}H(t_k)\Delta
t})=\hat{D}(\alpha(t))\exp(i\Phi(t))\label{propagatorDrivenOsc}
\end{equation}
where $\Delta t=t/n$, $t_k=k\Delta t$ and
\begin{eqnarray*}
\alpha(t)&=&\int_{0}^tdt^\prime\gamma(t^\prime), \\
\Phi(t)&=&\mbox{Im}\int_{0}^tdt^\prime\gamma(t^\prime)\int_{0}^{t^\prime}dt^{\prime\prime}\gamma^\ast(t^{\prime\prime}).
\end{eqnarray*}
In case of a driving force with constant amplitude,
$\gamma(t)=\Omega e^{i\delta t}$, one obtains
$\alpha(t)=i\left(\frac{\Omega}{\delta}\right)(1-e^{i\delta t})$
and $\Phi=\left(\frac{\Omega}{\delta}\right)^2(\delta t-\sin\delta
t)$. After a time $\tau_N=2\pi N/|\delta|$, $N=1,2,\ldots$, the
coherent state returns to its initial state in phase space with
its phase changed by an amount $\Phi(\tau_N)=2\pi
N\left(\frac{\Omega}{\delta}\right)^2\mbox{sign}(\delta)$. By
making this phase change depend on the internal states of a pair
of ions, an entangling gate operation can be achieved. For
\begin{equation}
H=i\hbar(\gamma(t)a^\dagger-\gamma^\ast(t)a){\cal O}
\label{DrivenOscillator}
\end{equation}
where ${\cal O}$ is an operator acting on the qubit states, the
propagator (\ref{propagatorDrivenOsc}) is replaced by
\begin{equation}
U_\gamma(t)=\hat{D}(\alpha(t){\cal O})\exp(i\Phi(t){\cal O}^2).
\label{propagatorHarmOsc}
\end{equation}
Choosing the interaction time $\tau$ such that $\alpha(\tau)=0$
thus realizes a propagator that depends nonlinearly on ${\cal O}$
and does not alter the vibrational state.

\subsection{Laser-ion interaction}
The interaction of a single ion qubit resonantly excited by a
monochromatic laser field with frequency $\omega_L$ is usually
described by performing a rotating-wave approximation with respect
to the optical frequency to obtain the Hamiltonian
\begin{equation}
H = \hbar\Omega\sigma_+ e^{-i\delta t}e^{i\eta(ae^{-i\nu
t}+a^\dagger e^{i\nu t})} +
\mbox{h.c.}\label{laserionHamiltonian},
\end{equation}
Here, $\delta=\omega_L-\omega_0$ is the detuning from the qubit
transition frequency $\omega_0$, $\nu$ is the frequency of the
ion's vibrational mode of interest, and
$\sigma_+=(\sigma_x+i\sigma_y)/2$ with the Pauli matrices
$\sigma_{x,y}$. The strength of the laser-ion coupling is
characterized by the Rabi frequency $\Omega$, and the strength of
processes involving changes in the vibrational state is determined
by the value of the Lamb-Dicke parameter $\eta$. Equation
(\ref{laserionHamiltonian}) represents a Hamiltonian in an
interaction picture that is defined with respect to the
Hamiltonian $H_0=\hbar\nu a^\dagger a+
\frac{\hbar\omega_0}{2}\sigma_z$ describing the ion qubit in the
absence of any laser-ion interactions. If $\eta\ll 1$, the
Lamb-Dicke approximation $e^{i\eta(ae^{-i\nu t}+a^\dagger e^{i\nu
t})}\approx 1+i\eta(ae^{-i\nu t}+a^\dagger e^{i\nu t})$ is used to
simplify (\ref{laserionHamiltonian}). The resulting three terms
describe excitations on the carrier, the lower and the upper
motional sideband, respectively. The generalization of the
Hamiltonian to the case of two and more ions is straightforward.
For the sake of simplicity, calculations in section
\ref{sec:EffectiveHamiltonian} will be limited to the case of the
laser coupling to the centre-of-mass mode along the axis of the
ion string where all ions experience the same coupling strength. A
detailed account of laser-ion interactions is given in
\cite{Leibfried03RMP}.

\subsection{$\sigma_z\otimes\sigma_z$ gate\label{subsec:zzGates}}
\begin{figure}
\centering
\includegraphics[width=11cm]{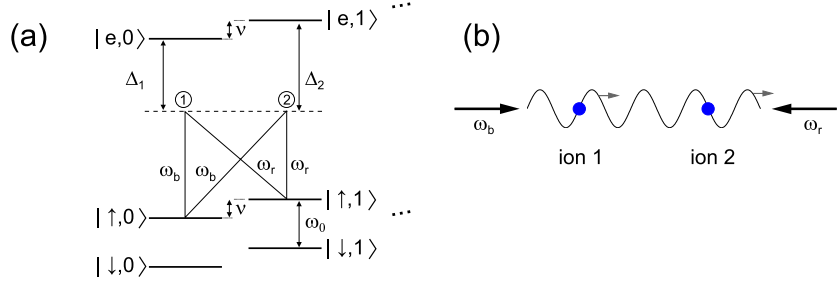}
\caption{Raman coupling of motional states for a hyperfine or
Zeeman qubit. (a) The states
$\ket{\uparrow,n=0},\ket{\uparrow,n=1}$ are coupled via the
excited state $\ket{\,e}$. A similar coupling not shown in the
figure exists for the qubit state $\ket{\downarrow}$. The laser
detunings $\Delta_1,\Delta_2$ from the mediating states are large
compared with trap frequency $\nu$ and the qubit level spacing
$\omega_0$ to avoid spontaneous emission from state $\ket{\,e}$.
Constructive interference of paths 1 and 2 is achieved for
counter-propagating laser beams. (b) The coupling is maximized for
counter-propagating laser beams with frequencies
$\omega_b$,~$\omega_r$ forming a moving standing wave along the
axis of vibration of the motional mode of
interest.}\label{fig1:MotionalCoupling_hyperfine}
\end{figure}

A Hamiltonian as described by (\ref{DrivenOscillator}) was
employed for the first time in an experiment \cite{Monroe96}
creating a Schr\"odinger cat state with a single ion using ${\cal
O}=\sigma_z$, i.e. a coupling to the motional mode that depended
on the internal energy eigenstate of the ion. Later, it was
realized that the same type of coupling could be used to entangle
a pair of ions by performing a conditional phase gate
\cite{Milburn00,Leibfried03}. In the experimental realizations
\cite{Leibfried03,Home06}, spin-dependent forces acting on a pair
of hyperfine or Zeeman ground states have been realized by near
resonant driving of Raman transitions between vibrational states
(see Figure~\ref{fig1:MotionalCoupling_hyperfine}). For this
purpose, two non-copropagating laser beams with frequencies
$\omega_b,\omega_r$ form a moving standing wave with difference
frequency $\omega_b-\omega_r$ close to the frequency $\nu$ of a
vibrational mode. The ac-Stark shift of the qubit states
$\ket{\downarrow}$,$\ket{\uparrow}$ results from a non-resonant
coupling to another atomic state $\ket{\,e}$  that is made qubit
state-dependent by properly chosen polarizations. Since the laser
field exhibits a strong spatiotemporal modulation, the resulting
potential gradients induce a force acting on the qubits that is
state-dependent and that couples to the vibrational mode by
displacing the qubit along a circle in phase space. When the ions
couple to the centre-of-mass mode (stretch mode), the coupling to
the mode can be made to disappear when both qubits are in the same
quantum state by choosing an ion spacing that is an odd (even)
integer multiple of half the wavelength of the moving standing
wave. Disregarding unimportant global and single qubit phases,
this coupling is then described by an operator ${\cal O}=S_z$
where $S_z=\sigma_z^{(1)}+\sigma_z^{(2)}$ is a collective spin
component of the qubits.

\begin{figure}
\centering
\includegraphics[width=12cm]{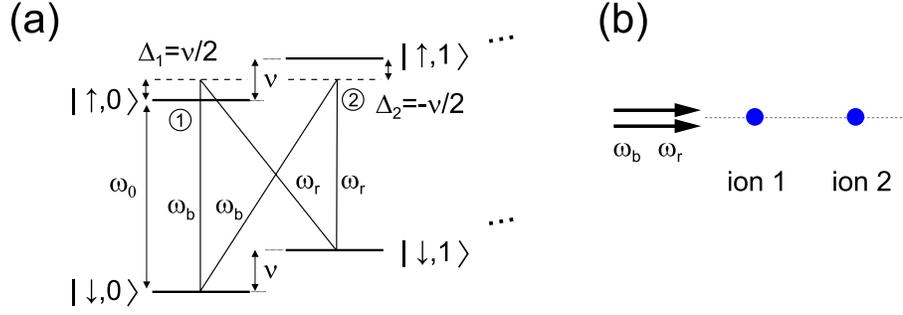}
\caption{Raman coupling of motional states for an optical qubit.
(a) The coupling for states
$\ket{\uparrow,n=0},\ket{\uparrow,n=1}$ is mediated by the states
$\ket{\downarrow,n=0},\ket{\downarrow,n=1}$ and vice versa on the
narrow qubit transition. Since spontaneous scattering from the
mediating state is small, the detuning can be made small compared
to the trap frequency. For $\Delta_1=\nu/2,\Delta_2=-\nu/2$, the
coupling is maximized by choosing a copropagating beam geometry.
(b) Optimum coupling is achieved for co-propagating laser beams
with frequencies $\omega_b$, $\omega_r$ propagating along the axis
of vibration of the motional mode of interest. As the bichromatic
laser field could also be described by a monochromatic laser field
that is amplitude-modulated with a frequency $\omega_b-\omega_r$
close to the vibrational frequency, no spatially varying ac-Stark
shifts are involved in the coupling.
}\label{fig1:MotionalCoupling_optical}
\end{figure}

The situation is different for qubits encoded in atomic states
connected by a narrow optical transition. For coupling motional
states $\ket{\downarrow,n=0}\leftrightarrow\ket{\downarrow,n=1}$,
here, the other qubit state $\ket{\uparrow}$ serves to mediate the
coupling. Similarly, a coupling between the states
$\ket{\uparrow,n=0}\leftrightarrow\ket{\uparrow,n=1}$ is mediated
by the state $\ket{\downarrow}$. To achieve a strong coupling, a
detuning $\Delta_i$ from the intermediate states can be chosen
that is smaller than the transition frequency between vibrational
states provided that the decay rate of the metastable state is
small compared to $\nu$. For $\omega_{b,r}=\omega_0\pm\nu/2$, the
two interfering paths shown in
Figure~\ref{fig1:MotionalCoupling_optical}a connecting levels
$\ket{\downarrow,n=0}\leftrightarrow\ket{\downarrow,n=1}$ have
equal strength. Since the detunings from the mediating states now
have opposite signs, destructive interference is achieved for
counter-propagating beams whereas the coupling is maximized for
co-propagating beams. In the limit of small excitation
($\Omega\ll\nu$), the coupling strength $\Omega_{R,0}$ on the
Raman transition between $\ket{\downarrow,n=0}$ and
$\ket{\downarrow,n=1}$ is given by
$\Omega_{R,0}=2\eta\Omega^2/\nu$. The states $\ket{\uparrow,n=0}$
and $\ket{\uparrow,n=1}$ are coupled with equal strength but
opposite sign. For stronger excitation, the carrier transition is
non-resonantly excited which leads to a saturation of
$\Omega_{R,0}$. As long as the intensities of the bichromatic
beams are equal, there is no overall ac-Stark shift due to
excitation of the carrier transition and the first motional
sidebands because ac-Stark shifts caused by the two laser fields
exactly cancel each other.

In the case of two ions excited on the centre-of-mass mode, a
driven quantum mechanical oscillator is realized with collective
atomic oscillator ${\cal O}=S_z$. In addition to the coupling of
vibrational states, there is another small M{\o}lmer-S{\o}rensen
coupling \cite{SorensenMolmer99} that does not exist for the case
of hyperfine or Zeeman qubits: collective spin flips between the
states $\ket{\downarrow\downarrow,n}$ and
$\ket{\uparrow\uparrow,n}$ occur by processes involving a blue and
a red photon that are mediated by the states
$\ket{\downarrow\uparrow,n\pm 1}$ and
$\ket{\uparrow\downarrow,n\pm 1}$ (see
Figure~\ref{fig:MSgate_scheme}). A similar process involving
either two blue or two red photons couples the states
$\ket{\uparrow\downarrow,n}$ and $\ket{\downarrow\uparrow,n}$.
\subsection{$\sigma_\phi\otimes\sigma_\phi$ gate\label{subsec:MSGates}}
\begin{figure}
\centering
\includegraphics[width=8cm]{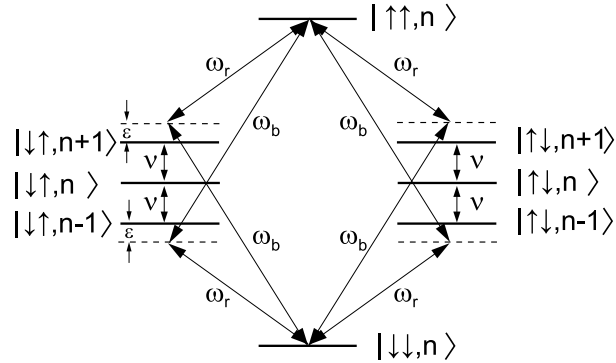}
\caption{M{\o}lmer-S{\o}rensen gate. A bichromatic laser field
with frequencies $\omega_b,\omega_r$ satisfying
$2\omega_0=\omega_b+\omega_r$ is tuned close to the upper and
lower motional sideband of the qubit transition. The field couples
the qubit states
$|\downarrow\downarrow\rangle\leftrightarrow|\uparrow\uparrow\rangle$
via the four interfering paths shown in the figure. Similar
processes couple the states
$|\uparrow\downarrow\rangle\leftrightarrow|\downarrow\uparrow\rangle$
with the same strength provided that the Rabi frequencies of the
light fields $\omega_b,\omega_r$ are
equal.}\label{fig:MSgate_scheme}
\end{figure}

%
In contrast to $\sigma_z\otimes\sigma_z$ gates that do not change
the internal states of the ions, the
$\sigma_\phi\otimes\sigma_\phi$ gate operations first investigated
by  A.~S{\o}rensen, K.~M{\o}lmer \cite{SorensenMolmer99} and
others \cite{Solano99} relies on collective spin flips
$\ket{\downarrow\downarrow}\leftrightarrow\ket{\uparrow\uparrow}$,
$\ket{\downarrow\uparrow}\leftrightarrow\ket{\downarrow\uparrow}$
by processes coupling to the lower and upper motional sidebands as
illustrated in Figure~\ref{fig:MSgate_scheme}.
It can be shown that the Hamiltonian governing the action of the
gate is described by setting ${\cal O}=\cos\phi S_x+\sin\phi S_y$
\cite{Wallentowitz95, Solano01}. For a properly chosen coupling
strength, the gate operations maps the product state basis
$\{\ket{\uparrow\uparrow}, \ket{\uparrow\downarrow},
\ket{\downarrow\uparrow}, \ket{\downarrow\downarrow}\}$ onto a
basis of entangled states. For hyperfine qubits, a detailed
discussion of advantageous beam geometries is presented in
\cite{Lee05}. In the case of optical qubits, it is again possible
to choose a pair of co-propagating beams for performing the gate
operation. The only difference to the $\sigma_z\otimes\sigma_z$
gate consists in the choice of laser frequencies
$\omega_{b,r}=\omega_0\pm\nu$ required for achieving a resonant
coupling. Formally, the gate operation is equivalent to a
$\sigma_z\otimes\sigma_z$ interaction in a rotated basis. To
stress this analogy, the M{\o}lmer-S{\o}rensen gate operation is
often also called a $\sigma_\phi\otimes\sigma_\phi$ gate.

The possibility of choosing co-propagating laser beams for
performing either $\sigma_z\otimes\sigma_z$ or
$\sigma_\phi\otimes\sigma_\phi$ gates is attractive from an
experimental point of view. The light field could be generated by
passing a laser beam through an acousto-optical modulator driven
by two radio-frequency fields and subsequently coupling the
first-order diffracted beams into a single-mode optical fibre,
thus realizing a simple and stable setup \footnote{In principle,
$\sigma_\phi\otimes\sigma_\phi$ gate operations could also be
achieved by phase-modulating a laser with a modulation frequency
close to $\nu$ by choosing a modulation index where the carrier
strength vanishes. However, this approach has the strong
disadvantage that even small changes of the modulation index from
the desired value give rise to light resonant with the transition
which is having desastrous effects on the gate performance.
Similarly, modulating at frequency close to $2\nu$ and using the
carrier and one of the first sidebands is problematic because of
light shifts induced by the other sidebands.}. If the Rabi
frequencies $\Omega_b$,~$\Omega_r$ of the blue and the red detuned
laser beam are equal, light shifts due to the non-resonant
excitation of the carrier transition and the first-order sidebands
are exactly cancelled. Light shifts arising from coupling to other
Zeeman transitions or far-detuned dipole transitions could be
cancelled by suitably balancing the ratio $\Omega_b/\Omega_r$.
\section{Effective Hamiltonians for $\sigma_z\otimes\sigma_z$
and $\sigma_\phi\otimes\sigma_\phi$ gates
\label{sec:EffectiveHamiltonian}}
We are interested in deriving an effective Hamiltonian that
accurately describes the dynamics on a pair of optical qubits
induced by a co-propagating bichromatic laser field with
frequencies $\omega_{b,r}=\omega_0\pm\delta$, where the detuning
is either close to half the vibrational frequency or close to the
vibrational frequency, i.~e. $\delta=(\nu-\epsilon)/2$ or
$\delta=\nu-\epsilon$ with $\epsilon\ll\nu$. As optical qubits
interacting with lasers typically have smaller Lamb-Dicke
parameters than hyperfine qubits coupled by Raman transitions,
transient non-resonant excitation of the carrier transition is
expected to play an important role for $\Omega\lesssim\nu$.
Whereas usually non-resonant interactions are taken into account
only qualitatively after having derived an effective Hamiltonian,
in the following calculation they will be eliminated right at the
beginning by going into a reference frame rotating at non-uniform
speed.

The Hamiltonian for the bichromatic laser field we are interested
in, is given by
\begin{equation}
H = \hbar\Omega e^{-i\phi}S_+ (e^{-i(\delta t+\zeta)}+e^{i(\delta
t+\zeta)})e^{i\eta(ae^{-i\nu t}+a^\dagger e^{i\nu t})} +
\mbox{h.c.}\,. \label{bichrHam}
\end{equation}
The laser is assumed to interact collectively with $m$ ions on the
axial centre-of-mass mode. Here,
$S_+=\sum_{i^=1}^m\sigma_+^{(i)}$, and $a$,~$a^\dagger$ denote
operators annihilating and creating phonons. It is also possible
to interpret this interaction as being due to a single resonant
laser beam that is amplitude-modulated with modulation frequency
$\delta$. The optical phase of the laser field is denoted $\phi$,
and the phase $\zeta$ accounts for a time difference between the
start of the gate operation and the maximum of the amplitude
modulation on the laser beam. Using the picture of an
amplitude-modulated resonant beam, it is obvious that there are
fast dynamical processes on the carrier transition with a
periodicity given by $\tau=2\pi/\delta$ that excite the ions to
the other state in the first half of the period and transfer it
back to the original state in the second half. We are not really
interested in exactly calculating the dynamical evolution of the
quantum state on this fast time scale. Rather, we would like to
know the time evolution at the instances $\tau, 2\tau,
3\tau,\ldots, \tau=2\pi/\delta$. It is useful to rewrite
(\ref{bichrHam}) as
\begin{eqnarray}
\fl H=
\hbar f(t)(e^{-i\phi}S_+\hat{D}(i\eta
e^{i\nu t})+e^{i\phi}S_-\hat{D}(-i\eta e^{i\nu t}))\nonumber\\
\hspace{-2.05cm}=
\hbar f(t)((S_x\cos\phi\!+\!S_y\sin\phi)(D_++D_-)+i(S_y\cos\phi\!-\!S_x\sin\phi)(D_+-D_-))\nonumber\\
\hspace{-2.05cm}=:
f(t)(S_x^{(\phi)}(D_++D_-)+iS_y^{(\phi)}(D_+-D_-))\,,\label{yetanotherHamiltonian}\nonumber
\end{eqnarray}
 where we used the displacement operator
$\hat{D}(\alpha)=e^{\alpha a^\dagger-a^\ast a}$ and the
definitions $\hat{D}_\pm=D(\pm i\eta e^{i\nu t})/2$,
$S_x^{(\phi)}=S_x\cos\phi +S_y\sin\phi$,
$S_y^{(\phi)}=S_y\cos\phi-S_x\sin\phi$ and
$f(t)=2\Omega\cos(\delta t+\zeta)$. For $\delta=\nu/2$ or
$\delta=\nu$ as required by either $\sigma_z\otimes\sigma_z$ or
$\sigma_\phi\otimes\sigma_\phi$ interactions, the Hamiltonian
is periodic in time, i.~e. $H(t+\tau)=H(t)$ with period
$\tau=2\pi/\delta$. Note that we assume the two-photon couplings
to be strictly resonant for the moment. In the next step, we will
get rid of the fast non-resonant carrier oscillation by going into
another interaction picture defined by $H_0=\hbar
f(t)S_x^{(\phi)}$. Writing
\begin{eqnarray*}
H&=&\hbar f(t)S_x^{(\phi)} + H_1,\mbox{\quad with}\\
H_1&=&
\hbar f(t)(S_x^{(\phi)}(D_++D_--1)+iS_y^{(\phi)}(D_+-D_-))\,,
\end{eqnarray*}
we obtain the interaction Hamiltonian
\begin{eqnarray}
\fl
H_{I}=e^{iF(t)S_x^{(\phi)}}H_1e^{-iF(t)S_x^{(\phi)}}\nonumber\\
\hspace{-1.9cm}=
\hbar f(t)S_x^{(\phi)}(D_+\!+\!D_-\!-\!1)+\hbar
f(t)\left(\cos(2F(t))S_y^{(\phi)}-\sin(2F(t))S_z\right)i(D_+\!-\!D_-),\nonumber
\end{eqnarray}
where
\begin{equation}
F(t)=\frac{2\Omega}{\delta}(\sin(\delta
t+\zeta)-\sin\zeta).\label{Ft_function}
\end{equation}
Now, we can approximate the time evolution over the course of an
oscillation period by a Magnus expansion of the propagator
\cite{Magnus54} in order to obtain
\begin{equation}
\fl
 U_I(t)=\exp\left\{-\frac{i}{\hbar}\left(\int_0^tdt^\prime
H_{I}(t^\prime)-
\frac{i}{2\hbar}\int_0^tdt^\prime\int_0^{t^\prime}dt^{\prime\prime}[H_{I}(t^\prime),H_{I}(t^{\prime\prime})]+
\ldots\right)\right\}.\label{BakerCampbellHausdorffPropagator}
\end{equation}
From now on, the phase $\phi$ will be set to zero to simplify the
notation. The formulas in the remainder of section
\ref{sec:EffectiveHamiltonian} are easily generalized to the case
of arbitrary $\phi$ by making the replacements $S_x\rightarrow
S_x^{(\phi)}$ and $S_y\rightarrow S_y^{(\phi)}$.

In the following two subsections, effective Hamiltonians for the
$\sigma_z\otimes\sigma_z$ gate and the
$\sigma_\phi\otimes\sigma_\phi$ gate will be derived starting from
(\ref{BakerCampbellHausdorffPropagator}). We are interested in
obtaining Hamiltonians of the form given in
(\ref{DrivenOscillator}) that are valid in the regime
$\Omega\ll\nu$. In addition, the calculation is going to yield
correction terms for the case when $\Omega\ll\nu$ no longer
strictly holds and additional terms that do not commute with the
atomic operator ${\cal{O}}$ in (\ref{DrivenOscillator}). Towards
this aim, terms proportional to higher orders of the expansion
parameter $(\Omega/\delta)$ will be dropped. In the calculation,
Bessel functions $J_n(x)$ will be evaluated at $x=4\Omega/\delta$.
These functions will be kept till the end of the calculation and
expanded in $\Omega/\delta$ only for the final analysis.

\subsection{$\sigma_z\otimes\sigma_z$ gate\label{subsec:zzGates_calc}}
For the $\sigma_z\otimes\sigma_z$ gate, we set $\delta=\nu/2$. We
start by calculating the first term
$H_{eff}^{(I,1)}=\frac{1}{\tau}\int_0^\tau dt^\prime
H_{I}(t^\prime)$ appearing in the exponent of
(\ref{BakerCampbellHausdorffPropagator}).
Here, it is important to note that the integrand
$H_I(t)=\sum_{k=-\infty}^\infty H_{(k)} e^{ik\delta t}$ is a
periodic function of time with period $\tau=2\pi/\delta$ so that
all its Fourier components except the constant term $H_{(0)}$ will
average to zero when they are integrated over one period. The only
non-zero Fourier components of the function
$f(t)=\Omega(e^{i(\delta t+\zeta)}+e^{-i(\delta
t+\zeta)})=\sum_{n=-\infty}^\infty f_n e^{in\delta t}$
are $f_{+1}=\Omega e^{i\zeta}$ and $f_{-1}=\Omega e^{-i\zeta}$.
Since all non-zero Fourier components of $(D_+\!+\!D_-\!-\!1)$ are
even integer multiples of $\delta$, the $S_x$-term of $H_{I}$
averages to zero. Therefore, we obtain
\[
H_{eff}^{(I,1)}=\frac{\hbar}{\tau}\int_0^\tau dt\,
f(t)\left(\cos(2F(t))S_y-\sin(2F(t))S_z\right)i(D_+-D_-)\,.
\label{HamFirstOrder}
\]
In the Lamb-Dicke limit,
\begin{eqnarray*}
i(D_+-D_-)&\approx&
-\eta(ae^{-i2\delta t}+a^\dagger e^{i2\delta
t})=:\sum_{n=-\infty}^\infty d_n e^{in\delta t}\,,
\end{eqnarray*}
and the components $d_{+2}=-\eta a^\dagger$ and $d_{-2}=-\eta a$
are the only relevant ones. Finally, we have
\begin{eqnarray*}
\fl
 \cos(2F(t))S_y-\sin(2F(t))S_z&=&
\frac{1}{2}\left[e^{i2F(t)}(S_y+iS_z)+e^{-i2F(t)}(S_y-iS_z)\right]\\
&=& Ae^{i\frac{4\Omega}{\delta}\sin(\delta t+\zeta)}+A^\dagger
e^{-i\frac{4\Omega}{\delta}\sin(\delta t+\zeta)}\\
&=& \sum_{n=-\infty}^\infty (AJ_n(\frac{4\Omega}{\delta})+
A^\dagger J_n(-\frac{4\Omega}{\delta}))e^{in\zeta}e^{in\delta t}\\
&=:&\sum_{n=-\infty}^\infty a_n e^{in\delta t}\,,
\end{eqnarray*}
where $J_n$ is a Bessel function,
$A=\frac{1}{2}(S_y+iS_z)e^{-i\psi}$ with
\begin{equation}
\psi=\frac{4\Omega}{\delta}\sin\zeta\,,\label{defpsi}
\end{equation}
and $a_n=(AJ_n(\frac{4\Omega}{\delta})+A^\dagger
J_n(-\frac{4\Omega}{\delta}))e^{in\zeta}=(A+(-1)^nA^\dagger)J_n(\frac{4\Omega}{\delta})e^{in\zeta}$.
In the following, the argument $4\Omega/\delta$ of the Bessel
functions $J_n$ will often be dropped to keep the notation simple.
It is convenient to express $A\pm A^\dagger$ as
\begin{equation}
\eqalign{S_{y\!\,,\psi}:=S_y\cos\psi+S_z\sin\psi=A+A^\dagger\cr
S_{z\!\,,\psi}:=S_z\cos\psi-S_y\sin\psi=-i(A-A^\dagger).}
\label{RotatedOps}
\end{equation}
Note that the linear transformation (\ref{RotatedOps}) preserves
the usual Lie algebra commutation relations for the operators
$S_x,S_{y\!\,,\psi},S_{z\!\,,\psi}$. The four terms
$f_{+1}d_{-2}a_{+1}$,$f_{-1}d_{+2}a_{-1}$, and to a lesser degree
$f_{+1}d_{+2}a_{-3}$, $f_{-1}d_{-2}a_{+3}$, contribute to
$H_{eff}^{(I)}$. Evaluating
\[
f_{+1}d_{-2}a_{+1}=(\Omega e^{i\zeta})(-\eta a)(A-A^\dagger)
J_1(\frac{4\Omega}{\delta})e^{i\zeta}=-\eta\Omega J_1
e^{2i\zeta}iaS_{z\!\,,\psi}
\]
as well as the other terms, we arrive at the effective Hamiltonian
\[
H_{eff}^{(I,1)} =
i\hbar\eta\Omega(J_1+J_3)(e^{-2i\zeta}a^\dagger-e^{2i\zeta}a)S_{z\!\,,\psi}\,,
\]
where $\eta\Omega(J_1+J_3)\approx
(2\eta\Omega^2/\delta)(1-4\Omega^2/(3\delta^2))$. This Hamiltonian
describes a spin-dependent force that starts to saturate when the
Rabi frequency goes up. While the atomic operator ${\cal
O}=S_{z\!\,,\psi}$ coincides in the limit of weak excitation with
the operator $S_z$ obtained from second-order perturbation theory,
it depends on the phase $\zeta$ between the blue- and the
red-detuned laser beams in the limit of strong excitation. For the
periodic Hamiltonian $H_I(t)=\sum_{k=-\infty}^\infty H_k
e^{ik\delta t}\,,$ the second order contribution to the effective
Hamiltonian $H_{eff}^{(I)}$ is given by
\[
H_{eff}^{(I,2)}=\frac{1}{\hbar\delta}\sum_{m=1}^\infty\frac{1}{m}[H_{(m)},H_{(-m)}]\,.
\label{HeffII2}
\]
After evaluating the commutators $[H_{(1)},H_{(-1)}]$,
$[H_{(3)},H_{(-3)}]$, the effective Hamiltonian
\[
H_{eff}^{(I)}=i\hbar\eta\Omega(J_1+J_3)S_{z\!\,,\psi}(a^\dagger
e^{-2i\zeta}-ae^{2i\zeta})-\frac{4\hbar\eta^2\Omega^2}{3\delta}{J_0}^2
S_{y\!\,,\psi}^2
\]
is obtained (the contribution of the commutator
$[H_{(2)},H_{(-2)}]\propto(\eta\Omega)^2(\Omega/\delta)^6$ is
insignificant). If the detuning $\delta=(\nu-\epsilon)/2$ slightly
deviates from half the oscillation frequency $\nu$, the
Hamiltonian is given by
\begin{equation}
H_{eff}^{(I)}=i\hbar\eta\Omega(J_1+J_3)S_{z\!\,,\psi}(a^\dagger
e^{i(\epsilon t-2\zeta)}-ae^{-i(\epsilon
t-2\zeta)})-\frac{4\hbar\eta^2\Omega^2}{3\delta}{J_0}^2
S_{y\!\,,\psi}^2 \label{Heffzz}
\end{equation}
is obtained. The second order term $H_{eff}^{(I,2)}$ account for
collective spin flip processes caused by a M{\o}lmer-S{\o}rensen
interaction. If this interaction did not exist, the propagator
could be calculated in the same way as for the driven harmonic
oscillator described by (\ref{DrivenOscillator}). In the limit
$\Omega\ll\nu$, where
$\eta\Omega(J_1+J_3)=2\eta\Omega^2/\delta+\Or(\Omega^4)$, the time
evolution from t=0 to $t^\ast=2\pi/|\epsilon|$ would create a
mapping of quantum states $\phi(0)\rightarrow\phi(t^\ast)$
described by the operator
\[
U_I(t^\ast)=\exp(i\theta t^\ast S_{z\!\,,\psi}^2)
\]
with
\[
\theta t^\ast
=\frac{\pi}{2}\left(\frac{4\eta\Omega^2}{\epsilon\delta}\right)^2\mbox{sign}(\epsilon).
\label{thetatast}
\]
For $m=2$ ions, the operator $U_I(t^\ast)$ performs a conditional
phase gate if $\theta t^\ast=\pi/8$. For $\zeta=0$ and weak
excitation ($\Omega\ll\nu$), this requires setting the coupling
strength $\Omega=\Omega_c$ with
\begin{equation}
\Omega_c^2=\frac{|\epsilon|\delta}{8\eta} \label{Omegazz}
\end{equation}
In the limit where $\Omega\ll\nu$ no longer holds, saturation
effects reduce the geometric phase $\Phi$ picked up in the gate
operation. For $\Omega=\Omega_c$, we would now have
\begin{equation}
\theta t^\ast\approx\frac{\pi}{8}\left(1-\frac{2}{3\eta
N_t}\right)\,, \label{thetatausat}
\end{equation}
where $N_t=\nu/|\epsilon|$ counts the number of trap cycles during
the gate operation. For $\eta=0.1$ and a gate time of 100 trap
cycles, $\Phi$ is reduced by about $7\%$. The smaller the
Lamb-Dicke factor gets, the more important saturation effects
become for a given gate time. The M{\o}lmer-S{\o}rensen
interaction contributes a term to the propagator $U_I(t^\ast)$
which is now approximately described by
\begin{equation}
U_I(t^\ast)\approx\exp(i\theta t^\ast
S_{z\!\,,\psi}^2)\exp(i\kappa t^\ast S_{y\!\,,\psi}^2)
\label{approxPropagatorzz}
\end{equation}
with
\begin{equation}
\kappa t^\ast
=\frac{\pi}{2}\left(\frac{4\eta\Omega^2}{\epsilon\delta}\right)^2\frac{|\epsilon|\delta}{3\Omega^2}.
\label{kappatast}
\end{equation}
 For the ratio $\kappa/\theta$,
\begin{equation}
\left|\frac{\kappa}{\theta}\right|=\frac{8}{3}\eta.
\label{kappaovertheta}
\end{equation}
If $\eta\ll 1$, the contribution from the second term $\propto
S_{y\!\,,\psi}^2$ is comparatively small.

Up to now, we have disregarded the fact that the effective
Hamiltonian is valid only for times $T=\frac{2\pi}{\delta}N,
N=1,2,\ldots$ where $\delta=\frac{1}{2}(\nu-\epsilon)$. Therefore,
the gate time $T$ needs to fulfil $|\epsilon| T=2\pi$ as well as
$\delta T=2\pi N$, with integer $N$. Combining both conditions, we
find
\[
\epsilon = \frac{\nu}{2N+1},\;\;N\in\mathbb{N}
\]

In writing equation (\ref{approxPropagatorzz}), terms arising from
the non-vanishing commutator $[S_{z\!\,,\psi},S_{y\!\,,\psi}^2]$
were neglected. Using the abbreviations
$\Omega_m=\eta\Omega(J_1+J_3)$ and
$\Omega_{MS}=4\eta^2\Omega^2{J_0}^2/(3\delta)$, it is convenient
to rewrite $H_{eff}^{(I)}=H_A+H_B$ with
\begin{eqnarray*}
H_A&=&\hbar\Omega_m iS_{z\!\,,\psi}(a^\dagger e^{i(\epsilon
t-2\zeta)} - a e^{-i(\epsilon
t-2\zeta)})-\frac{\hbar\Omega_{MS}}{2}(S_x^2+S_{y\!\,,\psi}^2)\\
H_B&=&\frac{\hbar\Omega_{MS}}{2}(S_x^2-S_{y\!\,,\psi}^2),
\end{eqnarray*}
since $H_A$ and $H_B$ commute. The time evolution induced by $H_A$
is given by the propagator
\begin{equation}
U_A(t)=\hat{D}(\lambda(t)S_{z\!\,,\psi})\exp(i\Phi(t)S_{z\!\,,\psi}^2)\exp(i\frac{\Omega_{MS}t}{2}(S_x^2+S_{y\!\,,\psi}^2))\label{PropagatorZZ}
\end{equation}
with $\lambda(t)=-ie^{-i2\zeta}(\Omega_m/\epsilon)(e^{i\epsilon
t}-1)$ and $\Phi(t)=(\Omega_m/\epsilon)^2(\epsilon t-\sin(\epsilon
t))$, and for the interaction Hamiltonian $H_{I,B}=U_A^\dagger H_B
U_A$ one finds
\[
H_{I,B}=\frac{\hbar\Omega_{MS}}{2}(\hat{C}(4\lambda(t))(S_x^2-S_{y\!\,,\psi}^2)+\hat{S}(4\lambda(t))\{S_x,S_{y\!\,,\psi}\}).
\label{HeffBzz}
\]
Here, the displacement operator
$\hat{D}(\pm\alpha)=\hat{C}(\alpha)\pm i\hat{S}(\alpha)$ was
expressed by the real-valued operators $\hat{C}$ and $\hat{S}$.
For the special case $\zeta=0$ this is equivalent to
\[
H_{I,B}=\frac{\hbar\Omega_{MS}}{4}(\hat{D}(-4\lambda(t))S_+^2 +
\hat{D}(4\lambda(t))S_-^2). \label{HeffBzz2}
\]
The last expression shows that the interaction Hamiltonian
$H_{I,B}$ describes collective spin flips between the levels
$\ket{\downarrow\downarrow}$ and $\ket{\uparrow\uparrow}$ that go
along with displacements of the vibrational state. For a phase
gate operation, $\max(|4\lambda(t)|)\approx 2$. Minimum
uncertainty states of motion are not conserved by the interaction.

\subsection{M{\o}lmer-S{\o}rensen gate operation \label{subsec:MSGates_calc}}
The formalism developed so far can be employed to study the
M{\o}lmer-S{\o}rensen gate as Hamiltonian (\ref{bichrHam}) also
describes the bichromatic laser field of the M{\o}lmer-S{\o}rensen
gate. Since the laser frequencies are set close to the blue and
red sideband resonance, the only difference is that
$\delta=\nu-\epsilon$ instead of
$\delta=\frac{1}{2}(\nu-\epsilon)$, thus changing the values of
the Fourier components $d_n$ used to express $D_\pm$. Taking into
account the leading terms in first and second order for the
calculation of (\ref{BakerCampbellHausdorffPropagator}), one finds
the effective Hamiltonian
\begin{equation}
\fl
 \frac{H_{eff}^{(I)}}{\hbar}=-\eta\Omega(\bess{0} +\bess{2})
S_{y\!\,,\psi}(a^\dagger e^{i(\epsilon t-\zeta)}+ae^{-i(\epsilon
t-\zeta)})-\frac{\eta^2\Omega^2}{2\delta}\bess{0}^2S_{y\!\,,\psi}^2
+\frac{2\eta^2\Omega^2}{3\delta}{J_1}^2S_{z\!\,,\psi}^2
\label{HeffMS}
\end{equation}
instead of (\ref{Heffzz}). Integrating from $t=0$ to
$t^\ast=\frac{2\pi}{|\epsilon|}$ and neglecting commutators
involving $S_{z\!\,,\psi}$ in the Magnus expansion, yields the
propagator
\begin{eqnarray}
\fl
 U_I(t^\ast)\approx\exp \left\{it^\ast
\left(\left(\frac{\eta^2\Omega^2}{\epsilon}((\bess{0}+\bess{2})^2
+ \frac{\eta^2\Omega^2}{2\delta}\bess{0}^2\right)S_{y\!\,,\psi}^2
-
\frac{2\eta^2\Omega^2}{3\delta}{J_1}^2S_{z\!\,,\psi}^2\right)\right\}\nonumber
\\
\hspace{-1.35cm}= \exp(i\lambda t^\ast S_{y\!\,,\psi}^2)
\exp(-i\mu t^\ast S_{z\!\,,\psi}^2)\label{MSPropagator}
\end{eqnarray}
with
\begin{eqnarray}
\lambda t^\ast &=&
\pi\frac{2\eta^2\Omega^2}{\epsilon|\epsilon|}\left((\bess{0}+\bess{2})^2+\frac{\epsilon}{2\delta}\bess{0}^2\right)\label{lambdat}\\
\mu t^\ast &\approx&
\pi\frac{16\eta^2\Omega^4}{3|\epsilon|\delta^3}\label{mut}.
\end{eqnarray}
The contribution $\propto1/(2\delta)$ comes from the
counter-rotating term, the red laser coupling to the blue sideband
and vice versa. For weak excitation, we have
\begin{eqnarray*}
U_I(t^\ast) &=&
\exp\left\{i\pi\frac{2\eta^2\Omega^2}{\epsilon|\epsilon|}
S_{y\!\,,\psi}^2 \right\}
\end{eqnarray*}
For $m=2$ ions, an entangling gate operation is achieved by
setting $|\lambda t^\ast|=\frac{\pi}{8}$ which amounts to setting
the coupling strength $\Omega$ to
\begin{equation}
\Omega_c=\frac{|\epsilon|}{4\eta}. \label{OmegaMS}
\end{equation}
In the limit where $\Omega\ll\nu$ is no longer valid but where
$|\epsilon/\eta|\ll\nu$ still holds, we find for the correction
terms in (\ref{lambdat}) when keeping $\Omega=\Omega_c$
\begin{equation}
|\lambda t^\ast| \approx \frac{\pi}{8}\left(1 -\frac{1}{4(\eta
N_t)^2}
 -\frac{\mbox{sign}(\epsilon)}{2N_t}\right).\label{lambdatsat}
\end{equation}
 and for the ratio
\begin{equation}
\left|\frac{\mu}{\lambda}\right|\approx\frac{1}{6(\eta N_t)^2
N_t}. \label{lambdaovermu}
\end{equation}
For $\eta=0.1$ and a gate operation that is performed within 100
trap cycles, the correction terms to $\lambda t^\ast$ have a
relative strength of $0.25\%$ and $0.5\%$, respectively, and the
$S_{z\!\,,\psi}^2$ interaction is less that $10^{-4}$ of the
$S_{y\!\,,\psi}^2$ term. Therefore, the interaction is quite well
approximated by using the propagator (\ref{MSPropagator}) with
$\mu t^\ast$ set to zero. Then, one obtains for arbitrary $t$
\[
U_I(t)=\hat{D}(\alpha(t)S_{y\!\,,\psi})\exp\left(i(\lambda t
-\chi\sin(\epsilon t))S_{y\!\,,\psi}^2\right)
\]
where
\begin{eqnarray}
\alpha(t)&=&\frac{\eta\Omega}{\epsilon}(\bess{0}+\bess{2})e^{-i\zeta}(e^{i\epsilon t}-1)\label{alpha}\\
\lambda &=&
\frac{\eta^2\Omega^2}{\epsilon}\left((\bess{0}+\bess{2})^2+\frac{\epsilon}{2\delta}\bess{0}^2\right)\label{lambda}\\
\chi&=&\frac{\eta^2\Omega^2}{\epsilon^2}(\bess{0}+\bess{2})^2\label{chi}
\end{eqnarray}
In the reference frame of the original Hamiltonian
(\ref{bichrHam}), the laser-ion interaction is therefore well
described by the propagator
\begin{equation}
U(t)=\exp(-iF(t)S_x)\hat{D}(\alpha(t)S_{y\!\,,\psi})\exp\left(i(\lambda
t -\chi\sin(\epsilon
t))S_{y\!\,,\psi}^2\right)\,.\label{MSPropagator2}
\end{equation}
\begin{figure}
\centering
\includegraphics[width=14cm]{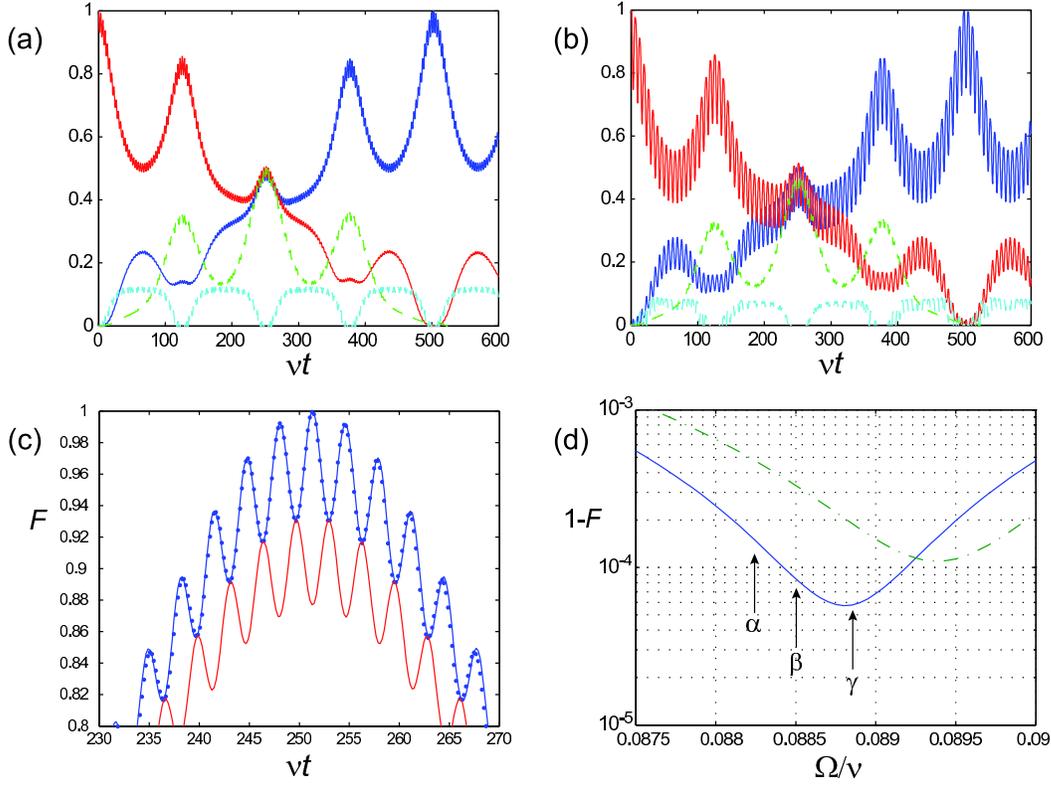}
\caption{(a) Time evolution of the density matrix elements of two
ions in a thermal state with $\overline{n\!}=2$ undergoing a
M{\o}lmer-S{\o}rensen interaction with $\Omega=0.0885\,\nu$,
$\eta=0.1$, $\epsilon=0.05\,\nu$ and $\zeta=0$. The calculations
are based on (\ref{MSPropagator2}). The values chosen reproduce
the curves shown in Figure~3(b) of reference
\cite{SorensenMolmer00}. Counting from above at $\nu t=60$, the
curves represent the populations
$\rho_{\downarrow\downarrow,\downarrow\downarrow}$,
$\rho_{\uparrow\uparrow,\uparrow\uparrow}$ and the coherences
$Im(\rho_{\downarrow\downarrow,\uparrow\uparrow})$ and
$Re(\rho_{\downarrow\downarrow,\uparrow\uparrow})$. At $\nu
t\approx 250$ ($t=4\pi/\epsilon$), the ions are in a maximally
entangled state. (b) Same as (a) but with $\zeta=\pi/2$. If the
gate operation starts in an intensity minimum of the
amplitude-modulated laser beam, the non-resonant carrier
oscillations are much stronger. At $\nu t=250$, the quantum state
is no longer maximally entangled. (c) Fidelity
$F=\bra{\psi_{max}}\rho(t)\ket{\,\psi_{max}}$ of creating the
maximally entangled state
$\ket{\psi_{max}}=(\ket{\downarrow\downarrow}-i\ket{\uparrow\uparrow})/\sqrt{2}$
near the optimum calculated from (\ref{MSPropagator2}). The upper
curve corresponds to $\zeta=0$, the lower one to $\zeta=\pi/2$.
The points on top of the upper curve represent the fidelity for
$\zeta=0$ and were obtained by a numerical integration of the
Hamiltonian (\ref{bichrHam}) after applying the Lamb-Dicke
approximation. (d) Infidelity $1-F$ of the gate at $\nu t=250$ for
$\zeta=0$ and a state with $\overline{n\!}=0$. The solid line is a
numerical integration of (\ref{bichrHam}) in the Lamb-Dicke
approximation, the dash-dotted line is based on the full
Hamiltonian. The arrow labelled '$\alpha$' denotes the optimum
Rabi frequency predicted by (\ref{OmegaMS}), '$\beta$' the value
of $\Omega$ chosen in \cite{SorensenMolmer00}, '$\gamma$' the Rabi
frequency predicted by
(\ref{lambda}).}\label{fig3:MS2000gate_comparison}
\end{figure}

This propagator can be used to calculate the dynamics of
expectation values of interest for the qubits. It is possible to
derive simple expressions by tracing over the motional states if
the vibrational mode is in a thermal state. For this, it is useful
to note that $\hat{D}(\alpha(t)S_{y\!\,,\psi})=\sum_\lambda
\hat{D}(\alpha(t)\lambda)P_\lambda$ where $P_\lambda$ denotes the
projector onto the subspace spanned by eigenvectors of
$S_{y\!\,,\psi}$ with eigenvalue $\lambda$. Moreover, the diagonal
elements of the displacement operator in the number-state
representation are given by $\langle n|\hat{D}(\alpha)|n\rangle =
\exp(-|\alpha|^2/2){\cal{L}}_n(|\alpha|^2)$ where ${\cal{L}}_n$
denotes a Laguerre polynomial \cite{Cahill69}. Since the
generating function of ${\cal{L}}_n(\beta)$ is given by
\cite{Abramowitz72}
\[
g(x,\beta)=\sum_{n=0}^\infty {\cal{L}}_n(\beta)
x^n=\frac{1}{1-x}\exp\left(-\frac{\beta x}{1-x}\right),
\]
summation over a thermal state with number state population
$p_n=\frac{1}{\bar{n}+1}(\frac{\bar{n}}{\bar{n}+1})^n$ and mean
phonon number $\bar{n}$ simply yields
\[
\sum_n p_n \langle
n|\hat{D}(\alpha)|n\rangle=\exp\left(-|\alpha|^2(\bar{n}+\frac{1}{2})\right).
\]
 In the case of two ions, $\zeta=0$, and an initial qubit state
$\rho_A=|\downarrow\downarrow\rangle\langle\downarrow\downarrow\!|$,
the expectation value $O(t)=\mbox{Tr}_Q({\cal O}\rho(t))$ of the
observable ${\cal O}$ is given by
\[
O(t)=\frac{1}{16}\mbox{Tr}_Q({\cal{O}_V}\{(S_z^2+S_x^2) -4S_z
e^{-4|\alpha|^2(\bar{n}+\frac{1}{2})}+(S_z^2-S_x^2)
e^{-16|\alpha|^2(\bar{n}+\frac{1}{2})}\})
\]
where $\mbox{Tr}_Q$ refers to the trace of the qubit state space
and ${\cal{O}_V}=V{\cal O}V^\dagger$ with
$V(t)=\exp(-iF(t)S_x)\exp(i\gamma(t) S_y^2)$ and $\gamma(t) =
\lambda t -\chi\sin(\epsilon t)$. As an example, the time
evolution of
$\langle\downarrow\downarrow|\rho(t)\ket{\downarrow\downarrow}$ is
explicitely given by
\[
\fl
p_{\downarrow\downarrow}(t)=\frac{1}{8}(2+\cos^2(2F))+\frac{1}{2}\cos(2F)\cos(4\gamma)e^{-4|\alpha|^2(\bar{n}+\frac{1}{2})}
+\frac{1}{8}\cos^2(2F)e^{-16|\alpha|^2(\bar{n}+\frac{1}{2})}
\]
with $\alpha(t)$,$\gamma(t)$,$F(t)$ containing the time dependent
terms. Other quantities of interest could be calculated in the
same way.

A propagator similar to (\ref{MSPropagator2}) was calculated in
ref.~\cite{SorensenMolmer00} for the case $\zeta=0$. The authors
argued that the non-resonant excitation of the carrier transition
could be neglected in a first step and obtained in this way a
Hamiltonian of the type described by (\ref{DrivenOscillator}) that
could be integrated exactly. In a second step, they considered the
influence of the previously neglected non-resonant excitations.
While this treatment yields correct results for $\zeta=0$, it
fails to predict the dependence of the gate operation on $\zeta$
via the angle $\psi$ as given by (\ref{MSPropagator2}).
Figure~\ref{fig3:MS2000gate_comparison} shows the time evolution
of matrix elements for the same parameters as used in
\cite{SorensenMolmer00} for the cases $\zeta=0$ and $\zeta=\pi/2$.
In the latter case, the amplitude of the non-resonant carrier
oscillations is considerable and the input state
$|\downarrow\downarrow\rangle$ is never perfectly mapped to a
maximally entangled state. For $\zeta\neq 0$, the effect of a
non-zero value of $\psi$ is fairly small for current gate
realizations using hyperfine qubits where the Lamb-Dicke parameter
$\eta$ is considerable. However, it becomes crucial for the
realization of fast gates on optical qubits with small $\eta$
since in this case the gate requires a larger value of $\Omega$ to
achieve the same gate speed.
\begin{figure}
\centering
\includegraphics[width=8cm]{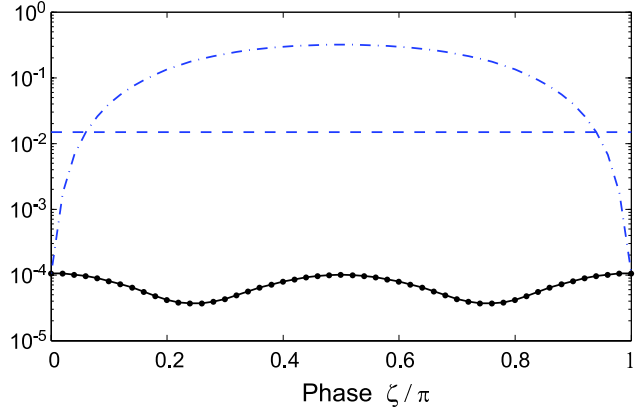}
\caption{Performance of the $\sigma_\phi\otimes\sigma_\phi$ gate
as a function of the phase $\zeta$ for a gate operation taking
place in 25 trap cycles with $\eta=0.05$, $\delta=0.96\,\nu$ and
$\Omega=0.221\,\nu$. The figure shows the distance
$d(U_{ex},U_{id,\phi})$ between the exact propagator $U_{ex}$
obtained by numerical integration of (5)
 and the
gate operation $U_{id,\phi}=\exp(i\frac{\pi}{8}S_{y\!\,,\psi}^2)$
(solid line). The points on top of the line denote
$d(U_{ex},U_{(27)})$ where
$U_{(27)}$ is the propagator given by
(27), thus demonstrating that this equation is a
very good approximation to the exact solution. The dashed-dotted
line shows $d(U_{ex},U_{id,\phi})$ with
$U_{id,\phi}=\exp(i\frac{\pi}{8}S_{y}^2)$ and the dotted line
$d(U_{ex},U_{pert})$ with
$U_{pert}=\exp(i2\pi\eta^2\Omega^2/\epsilon^2S_{y}^2)$ as
predicted by simple second-order perturbation theory. More details
regarding the distance measure $d$ are given in the text.
\label{fig5:CompareMolmergate}}
\end{figure}

Figure~\ref{fig5:CompareMolmergate} shows a comparison of the
different propagators for the gate operation with $\eta=0.05$
taking place in 25 trap cycles ($\delta=0.96\nu$). For the
prediction of the required coupling strength $\Omega$ for a gate
operation realizing
$U_{id,\phi}=\exp(i\frac{\pi}{8}S_{y\!\,,\psi}^2)$, equation
(\ref{lambda}) was iteratively solved to yield $\lambda
t^\ast=\pi/8$. The propagator $U_{ex}$ was obtained from a
numerical integration of (\ref{bichrHam}). Then, $U_{ex}$ was
compared to $U_{id,\phi}$, to the propagator of
(\ref{MSPropagator2}), to $U_{id}=\exp(i\frac{\pi}{8}S_{y}^2)$ and
to the prediction
$U_{pert}=\exp(i2\pi\eta^2\Omega^2/\epsilon^2S_{y}^2)$ of
second-order perturbation theory. Since the exact propagator does
not perfectly return the motional state to the initial state at
the end of the gate, the following procedure was applied for the
calculation of the distance $d$ between the propagators: we assume
that the ions are initially in the motional ground state and that
a cooling mechanisms returns the motional state to the ground
state at the end of the gate operation without affecting the qubit
states. This turns the unitary evolution into a quantum process
acting only on the internal states of the ions. For the comparison
of two quantum processes $\mathcal{E}_1$, $\mathcal{E}_2$, the
processes are mapped using the Jamiolkowski isomorphism onto
density matrices $\rho_1$, $\rho_2$ for which the distance
$d(\rho_1,\rho_2)=1-\tr(\sqrt{\sqrt{\rho_1}\rho_2\sqrt{\rho_1}})$
is calculated \cite{Gilchrist05,Duer05}. The results show that the
propagator given by (\ref{MSPropagator2}) correctly predicts the
coupling strength as well as the operator realized by the gate
operation. It also becomes obvious that $U_{id}$ considerably
deviates from the operation generated by the Hamiltonian
(\ref{bichrHam}) unless $\zeta=0$.

\subsection{Comparison of $\sigma_z\otimes\sigma_z$ and $\sigma_\phi\otimes\sigma_\phi$ gates
\label{subsec:Comparison}}
The main advantage of the $\sigma_z\otimes\sigma_z$ interaction on
optical qubits appears to be its insensitivity to changes in the
optical path length. In the limit of weak excitation, the gate
operation tolerates changes that occur within the gate operation
as in each elementary process a photon is absorbed and another one
emitted into the same laser beam (this property does not hold for
hyperfine qubits since here Raman beams in a counter-propagating
configuration are used). If higher Rabi frequencies are used, the
interaction rather becomes
$\sigma_{z\!\,,\psi}\otimes\sigma_{z\!\,,\psi}$ which make it
susceptible to path-length fluctuations within the gate time.
Still, if amplitude-shaped pulses are applied (see section
\ref{sec:ShapedPulses}), the gate operation tolerates changes of
the path length that occur between consecutive applications of the
gate. This is not the case for the M{\o}lmer-S{\o}rensen gate
which becomes robust against changes between gate operations but
remains susceptible to changes occurring within the gate when the
gate is sandwiched between $\pi/2$ pulses applied to both qubits.
\begin{table}
\centering
\begin{tabular}{|c||c|c|}
  \hline
  Gate type & $\sigma_z\otimes\sigma_z$ & $\sigma_\phi\otimes\sigma_\phi$ \\
  \hline\hline
  Rabi frequency & $\Omega/\nu=1/(4\sqrt{\eta N_t})$ & $\Omega/\nu=1/(4\eta N_t)$ \\
  \hline
  Saturation strength & $\gamma=2/(3\eta N_t)$ & $\gamma=1/(4\eta^2 N_t^2)$ \\
  \hline
  Coupling ratio & $\kappa/\theta=8/(3\eta)$ & $\mu/\lambda=1/(6\eta^2 N_t^3)$ \\
  \hline
\end{tabular}
 \caption{Comparison of $\sigma_z\otimes\sigma_z$
and $\sigma_\phi\otimes\sigma_\phi$ gates. The first row gives the
Rabi frequency $\Omega$ required to peform an entangling gate
operation as a function of $\eta$ and the gate duration. The
latter is expressed as the number of trap oscillation periods
$N_t$. The second row lists the reduction in coupling strength for
this kind of gate due to saturation effects. The third row
compares the unwanted to the desired coupling
strength.\label{GateCompTable}}
\end{table}

The $\sigma_z\otimes\sigma_z$ gate, however, seems to be much less
favorable with respect to the following criteria: (i) the Rabi
frequency that is required for performing the gate operation in a
given time, (ii) the strength of saturation effects reducing the
coupling for the Rabi frequency needed for the gate operation, and
(iii) the ratio between the desired and the unwanted spin-spin
couplings. Table \ref{GateCompTable} shows a comparison of the
gates with respect to these criteria, thus summarizing the results
(\ref{Omegazz}),~(\ref{thetatausat}),~(\ref{kappaovertheta}) for
the $\sigma_z\otimes\sigma_z$ gate and
(\ref{OmegaMS}),~(\ref{lambdatsat}),~(\ref{lambdaovermu}) for the
$\sigma_\phi\otimes\sigma_\phi$ gate. For all three criteria, the
M{\o}lmer-S{\o}rensen performs better. Having a low Rabi frequency
is also of interest when it comes to non-resonant excitation of
other vibrational modes or light shifts induced by excitation of
far-detuned dipole transitions.
\section{Amplitude-shaped pulses and spin echos \label{sec:ShapedPulses}}
\subsection{Amplitude-shaped laser pulses}
In the limit of fast gate operations, the Hamiltonians
(\ref{Heffzz}), (\ref{HeffMS}) become sensitive to the phase
$\zeta$ which is related to the intensity of the bichromatic laser
field at the start of the gate operation. It is therefore
interesting to shape the intensity of the bichromatic laser field
during the gate operation so that the atomic operator ${\cal
O}(t)=S_{j,\psi(t)}$, with
$\psi(t)=\frac{4\Omega(t)}{\delta}\sin\zeta$, appearing in the
Hamiltonians  becomes time-dependent but independent of $\zeta$ at
the beginning and at the end of the gate when the intensity is
low. In this way, the gate could be made insensitive to $\zeta$ by
an adiabatic process where ${\cal O}(t)$ evolves  from $S_{j}$ at
the start of the gate operations to a $\zeta$-dependent operator
$S_{j,\psi(t)}$ and back to $S_{j}$. However, the state
$\alpha(t)$ of the vibrational mode generally does not return to
its original state at the end of the gate under the action of the
propagator (\ref{propagatorHarmOsc}) when the coupling $\gamma(t)$
is made time-dependent. There is, however, a class of shaped
pulses with the property $\alpha(\tau)=0$ that can be constructed
in the following way: By applying an amplitude-shaped pulse twice
with a sign change in the coupling between the two pulses, i.~e.
$\gamma_2(t)=-\gamma_1(t)$, one obtains the propagator
\[
U=U_{-\gamma}(2\tau,\tau)U_\gamma(\tau,0)= \exp(i2\Phi(\tau){\cal
O}^2)
\]
because the first and the second pulse displace the motional state
into opposite directions but by an equal amount (see
(\ref{propagatorHarmOsc})). A quantum state that is displaced
along a circle in phase space by an off-resonant force of constant
magnitude, $\gamma(t)=\Omega e^{i\epsilon t}$,
$t\in[0,2\pi/\epsilon]$, can be viewed as a special case of this
pulse form with $\tau=\pi/\epsilon$. For the bichromatic gates
based on the interactions (\ref{Heffzz}), (\ref{HeffMS}), the sign
change can be accomplished by either shifting the phase $\zeta$
during the action of the second pulse by an amount $\pi/2$
($\pi$), respectively, or by changing the overall phase of the
laser by $\pi$ during the second pulse (i.e. $\Omega\rightarrow
-\Omega$).

\begin{figure}
\centering
\includegraphics[width=12cm]{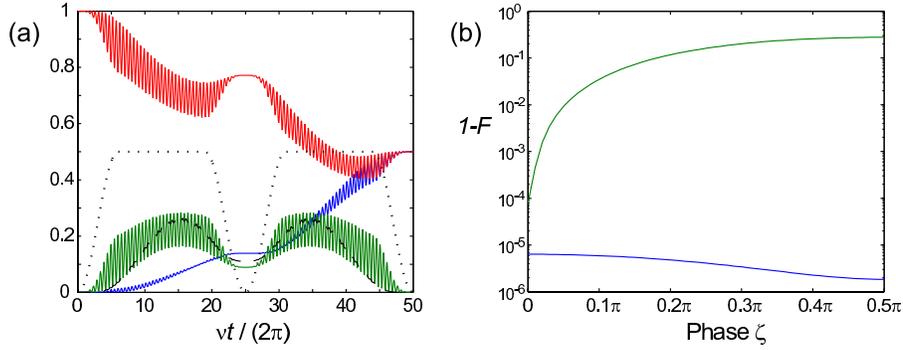}
\caption{M{\o}lmer-S{\o}rensen gate operation with two
amplitude-shaped laser pulses based on the full Hamiltonian
(\ref{bichrHam}) without Lamb-Dicke approximation. For parameters
$\Omega_{max}=\max(|\Omega|) = 0.167\,\nu$, $\eta=0.05$, the gate
takes place during 50 trap oscillation periods. The pulses are
switched on and off during 8 trap cycles using a $\cos^2$-profile.
During the second pulse, the phase of the blue- and the
red-detuned beam is shifted by $\pi$ with respect to the first
pulse. (a) Time evolution of the populations
$p_{\uparrow\uparrow},p_{\downarrow\downarrow},p_{\downarrow\uparrow}+p_{\uparrow\downarrow}$.
when starting from state $|\uparrow\uparrow,n=0\rangle$ at time
$t=0$. The dotted line shows the coupling strength
$|\Omega(t)|/(2\Omega_{max})$, the dashed line is the average
number of vibrational quanta. (b) Infidelity of the final state as
a function of the phase $\zeta$. The upper curve shows the strong
influence of the phase for a gate operation with constant coupling
strength $\Omega=0.147\,\nu$ where a high-fidelity operation is
achieved only for $\zeta=0$. For the amplitude-shaped gate, the
fidelity is practically independent of $\zeta$. Similar results
are also obtained for other input states. For a realistic
calculation of $F$, decoherence caused by spontaneous decay of the
metastable state would have to be taken into account.
}\label{fig4:ShapedPulses}
\end{figure}
Figure~\ref{fig4:ShapedPulses} illustrates the use of
amplitude-shaping in order to make the M{\o}lmer-S{\o}rensen gate
operation robust against fluctuations in the phase $\zeta$ between
the blue- and the red-detuned laser beam. In this example, an
entangling gate is accomplished within $N=50$ trap oscillation
periods by a pair of laser pulses with
$\Omega(t+\tau)=-\Omega(t)$, for $t\le\tau$ with $\tau=\pi N/\nu$.
As shown in Figure~\ref{fig4:ShapedPulses}a, the pulses are
switched on and off within eight trap cycles. After the first
pulse, the vibrational state has not returned to its initial
state. It is only after the second pulse that the correlations
between the vibrational state and the qubit states vanish again.
Using this technique, the initial state
$|\uparrow\uparrow,n=0\rangle$ is mapped to the target state
$\frac{1}{\sqrt{2}}(|\uparrow\uparrow\rangle+i|\downarrow\downarrow\rangle)|n=0\rangle$
with an infidelity of below $10^{-5}$ (see
Figure~\ref{fig4:ShapedPulses}b). This is in sharp contrast to the
case of an excitation of the same duration with constant amplitude
where the infidelity depends on the phase $\zeta$ and varies
between $10^{-4}$ and $0.2$. Similar results are also obtained for
other input states.
\subsection{Spin echos}
Spin echo pulses can be combined with amplitude-shaped laser
pulses to make the $\sigma_z\otimes\sigma_z$ gate more robust
against imperfections. It is possible to implement the conditional
phase gate operation by having the motional state perform two
circles in phase space so that the gate pulse can be split up into
two separate pulses. Since the quantum states
$\ket{\uparrow\uparrow}$, $\ket{\downarrow\downarrow}$ as well as
the states $\ket{\uparrow\downarrow}$, $\ket{\downarrow\uparrow}$
pick up the same phases
$\Phi_{\uparrow\uparrow}=\Phi_{\downarrow\downarrow}$,
$\Phi_{\uparrow\downarrow}=\Phi_{\downarrow\uparrow}$, it is
possible to exchange the states
$\ket{\uparrow\uparrow}\leftrightarrow
\ket{\downarrow\downarrow}$,
$\ket{\downarrow\uparrow}\leftrightarrow \ket{\uparrow\downarrow}$
by a collective $\pi$-pulse sandwiched between the two gate pulses
and to exchange the populations at the end of the gate sequence
again. The first spin-echo $\pi$-pulse inverts the direction of
the force on the motional state so that the motional state returns
to the initial state after the second spin-dependent pulse. In
contrast to the case of shaped pulses without spin-echo, there is
no need for changing the phase $\zeta$ of the second pulse or the
sign of the coupling strength $\Omega$. The spin echo procedure is
advantageous for the following reasons:
\begin{enumerate}
\item The gate becomes more robust against unequal light
intensities on the ions.
\item Single qubit phases arising from light shifts are
transformed into an unimportant global phase. In the context of
this gate, light shifts will mainly be due to an imbalance in the
power of the blue and the red-detuned laser beams and also due to
very off-resonant excitation of dipole transitions. In addition, a
light shift $\delta_{ls}$ occurs if the average frequency
$\omega_L$ of the bichromatic light field does not exactly
coincide with the atomic transition frequency $\omega_0$. However,
this light shift will be fairly small as
$\delta_{ls}\propto(\Omega/\delta)^2(\omega_L-\omega_0)$.
\item Collective spin flips arising from the term $S_y^{\,2}$ in
(\ref{Heffzz}) can be cancelled to first order by choosing
rotation axes for the $\pi$-pulses on ion 1 and ion 2 that differ
by $90^\circ$ (x-rotation on ion 1 and y-rotation on ion 2). This
effectively changes the sign of the rotation angle $\kappa$
occurring in (\ref{approxPropagatorzz}) for the second pulse and
eliminates the spin flip contribution of the interaction. To
perform different $\pi$-pulses on both ions requires, however,
either a different trap frequency that changes the distance
between the ions by $\lambda/4$ or an additional laser beam.
\end{enumerate}
In the limit of short gate operations, spin echos become somewhat
less efficient in cancelling perturbations described by $S_z$
interactions as the gate interaction $\propto S_{z,\psi}^2$ no
longer commutes with $S_z$ for $\psi\neq 0$.

For the M{\o}lmer-S{\o}rensen gate operation, where
$[S_{y,\psi}^2,S_z]$ is not a small quantity, spin echos seem to
be of limited use. If, however, the gate interaction is sandwiched
between a pair of collective $\pi/2$ pulses to turn it into a
$\sigma_z\otimes\sigma_z$ interaction, spin echos are helpful for
cancelling perturbations occurring between consecutive gates.
Also, it should be noted that a spin-echo like technique was
already proposed in ref. \cite{SorensenMolmer99} in order to cope
with number-state dependent ac-Stark shifts that arise if the gate
is implemented by illuminating ion 1 with a red-detuned laser beam
and ion 2 with a blue-detuned laser beam instead of using a
bichromatic light field for both ions.

\section{Conclusions}
Collective laser-ion interactions with bichromatic laser beams are
capable of performing both $\sigma_z\otimes\sigma_z$ gates as well
as M{\o}lmer-S{\o}rensen gate operations. The analysis shows that
it is important to include non-resonant excitation of the carrier
transition for the precise calculation of the gate operation.
While the paper was focused on the case of qubit states linked by
a weak optical transition, the discussion of the
M{\o}lmer-S{\o}rensen gate interaction applies also to hyperfine
qubits where non-resonant carrier excitation also occurs in the
limit of fast gate operations. For optical qubits, the required
laser beams can chosen to be co-propagating which allows for a
robust and experimentally easily realizable setup where an
acousto-optical modulator is used in single-pass configuration to
create the bichromatic light field. In a direct comparison of the
gates, the M{\o}lmer-S{\o}rensen interaction seems to be
advantageous in terms of required laser power and gate accuracy
while the $\sigma_z\otimes\sigma_z$ interaction has the advantage
of being robust even against certain path length fluctuations
occuring during the gate operation in the limit of weak driving
where $S_{z\!\,,\psi}\approx S_z$. For gate durations coming close
to $T=2\pi/(\eta\nu)$, control of the phase $\zeta$ between the
red- and the blue-detuned laser beams is of vital importance
unless the gate is performed using amplitude-shaped laser pulses.
In this case, the requirements are strongly relaxed and the gates
appear to be very promising for experimental realization. The
possibility of using a single laser beam for global single qubit
and entangling operations opens also interesting perspectives for
creating multi-particle entangled states with more than two ions.
The operations using this beam could be combined with an
off-resonant strongly focussed beam capable of inducing $\sigma_z$
operations on individual qubits in order to create a larger
variety of complex entangled states.
%
\ack
This work was supported by the Austrian Science Fund (FWF), DTO
and the EU network SCALA. The author wishes to thank W. D\"ur for
discussions and J. Benhelm for helpful comments and a critical
reading of the manuscript.

\end{document}